\documentclass[runningheads]{llncs}
\usepackage[T1]{fontenc}
\usepackage{graphicx}
\usepackage{cite}
\usepackage{amsmath,amssymb,amsfonts}
\usepackage{algorithmic}
\usepackage{graphicx}
\usepackage{textcomp}
\usepackage{xcolor}
\usepackage{pifont}
\newcommand{\cmark}{\ding{51}}%
\newcommand{\xmark}{\ding{55}}%
\usepackage{float}
\usepackage{nicematrix}
\usepackage{multirow}
\usepackage{booktabs}
\usepackage{subcaption}
\usepackage{float}
\usepackage[hidelinks]{hyperref}
\newcommand{\yt}{YouTube\ }
\usepackage{comment}
\usepackage{verbatim} 
\usepackage{adjustbox}
\usepackage{placeins} 
\usepackage{afterpage}
\usepackage{listings}
\begin{document}

\title{Half-life of Youtube News Videos: Diffusion Dynamics and Predictive Factors}

\author{
Anahit Sargsyan\inst{1} \and
Hriday Sankar Dutta\inst{2} \and
Jürgen Pfeffer\inst{1}
}

\institute{
\inst{1}School of Social Sciences and Technology, Technical University of Munich, Germany\\
\email{\{anahit.sargsyan, juergen.pfeffer\}@tum.de}
\and
\inst{2}Deakin University, GIFT City, India\\
\email{hridoy.dutta@deakin.edu.au}
}

\maketitle

\begin{abstract}
Consumption of \yt news videos significantly shapes public opinion and political narratives. While prior works have studied the longitudinal dissemination dynamics of \yt News videos across extended periods, limited attention has been paid to the short-term trends. In this paper, we investigate the early-stage diffusion patterns and dispersion rate of news videos on \yt, focusing on the first 24 hours. To this end, we introduce and analyze a rich dataset of over 50,000 videos across 75 countries and six continents. We provide the first quantitative evaluation of the 24-hour half-life of YouTube news videos as well as identify their distinct diffusion patterns. According to the findings, the average 24-hour half-life is approximately 7 hours, with substantial variance both within and across countries, ranging from as short as 2 hours to as long as 15 hours. Additionally, we explore the problem of predicting the latency of news videos' 24-hour half-lives. Leveraging the presented datasets, we train and contrast the performance of 6 different models based on statistical as well as Deep Learning techniques. The difference in prediction results across the models is traced and analyzed. Lastly, we investigate the importance of video- and channel-related predictors through Explainable AI (XAI) techniques. The dataset, analysis codebase and the trained models are released at \href{http://bit.ly/3ILvTLU}{\color{blue}http://bit.ly/3ILvTLU} to facilitate further research in this area.\looseness-2

\end{abstract}

\section{Introduction}

YouTube, since its launch in early 2005, has grown into one of the largest and most popular video-sharing platforms with nearly 10 billion public videos \cite{mcgrady2023dialing}. The platform allows creators to share their content and users to interact with the published videos through likes, dislikes, and comments. By 2021, the platform had reached 2.3 billion users globally \cite{Janki}, with no competitor achieving similar success \cite{richier2014modelling}. Apart from creating new forms of entertainment and marketing, YouTube has notably transformed media industries, from music \cite{cayari2011youtube} to news and television \cite{al2017boots}. Through its recommendation algorithms, YouTube has been shown to play an important role in shaping public knowledge, attitudes, and opinions on a global scale \cite{rieder2018ranking, bryant2020youtube, chen2021exposure}. 
Apart from the social aspect, the spread of YouTube videos has been demonstrated to have a major impact on political landscapes~\cite{  vesnic2014youtube, Lee_Wu_Ertugrul_Lin_Xie_2022, epstein2023youtube}. For instance, Epstein et al.~\cite{epstein2023youtube} quantified the impact of video recommendation on voting preferences in the US through randomized controlled trials with a YouTube simulator. The results show that the preferences of undecided voters shifted dramatically (between 51.5\% and 65.6\%) towards candidates featured in the videos. \looseness-2

By examining the dissemination dynamics of YouTube videos and their contributing factors, one can gain valuable insights into digital marketing strategies, the propagation of misinformation, the role of online media in societal change, to name a few. In \cite{Meeyoung}, Cha et al. studied the popularity life-cycle of Youtube videos randomly sampled from "Entertainment" and "Science and Technology" categories, and identified the key elements that shape the popularity distribution. The results revealed that the popularity distribution generally exhibits power-law behavior, yet the exact distribution seems category-dependent. Along the same lines, Figueiredo et al.~\cite{figueiredo2011tube} analyzed a dataset of over 100,000 YouTube videos to characterize the growth patterns of popularity since the time of upload. The dataset included “top”, “deleted” and “random” videos and the analysis revealed that bursts of popularity tend to be caused by external search traffic and referrals. Also, according to the findings, it takes 26\%, 5\% and 43\% of their total lifetimes to reach their half lives for videos in “top”, “deleted” and “random” datasets respectively. In \cite{Zeni}, Zeni et al. examined the diffusion dynamics of YouTube videos, focusing on videos from the Music and News categories. The analysis of over 54,000 news videos demonstrated that, unlike the Music category, these videos are relevant to the YouTube community only for a \textit{very short period} of time after publication. Taking a step further, Yu et al.~\cite{Yu_Xie_Sanner_2021} examined a temporal dataset containing the 2-year history of over 172,000 YouTube videos to study the long-term popularity trends among different types of videos, including news videos. According to the findings, more than 60\% of news videos are dominated by one long power-law decay (i.e., the number of views peaks within a short time-frame then decreases drastically over the course of a few days), confirming the observations reported in \cite{Zeni}. While these studies have provided insights into the longitudinal growth dynamics of YouTube News videos across extended periods, the short-term trends remain under-explored. 

To fill this gap, we investigate the early-stage distribution patterns of news videos on YouTube, focusing on the first 24 hours. The analysis covers a broad global perspective, spanning 75 countries, along with comparisons between specific countries. The dynamics is studied with respect to the so-called 24-hour \textit{half-life} of news videos (for brevity, hereafter referred to as half-life), which is the amount of time required for a video to reach half of the 24-hour views. Additionally, we train and evaluate various Machine Learning models for predicting the latency of news videos' half-lives. More concretely, the key contributions of the present work are four-fold:

\begin{itemize}
    \item First, we quantify the half-life of YouTube news videos
from 75 countries across six continents. The statistics of the included news channels are provided, and the variation in half-lives is discussed. According to the findings, half-life of YouTube news videos ranges from 2 to 15 hrs. with an average of 7.32. The shortest country-wise average half-life (of 4 hrs.) is observed in Malta and Taiwan, whereas the longest (of 15 hrs.) in Saudi Arabia.
    \item Second, we provide a localized fine-grained analysis investigating the diffusion patterns and consumption rate of YouTube news videos in the US and Germany. Four distinct diffusion patterns were identified and category-wise comparative evaluation of half-lives and news categories is presented. 
    \item Next, we explore the problem of predicting the latency of news videos' half-lives and shed light on the importance of selected video- and channel-related predictors. The problem is cast as a binary classification task (early vs late), and 6 different models, based on both statistical as well as Deep Learning methods, are evaluated. Through comparison, we arrive at an ensemble tree-based classifier that achieves an average F1-score of nearly $82$\%. The analysis of predictive factors through Explainable AI (XAI) techniques revealed that the most impactful features were number of videos posted by the channel, average half-life of the country, age of the channel and video length.  
    \item Lastly, we introduce two datasets containing the metadata and 24-hour view counts of YouTube news videos. The first dataset was compiled from 48,406 videos sourced from 624 prominent news channels across 75 countries, whereas the second was from over 10,500 videos from German and US YouTube news channels covering 71 channels. The \textit{database}, \textit{trained models}, and the complete source code to reproduce the conducted analysis can be accessed online at \href{http://bit.ly/3ILvTLU}{\color{blue}http://bit.ly/3ILvTLU}.
\end{itemize}

The remainder of this paper is organized as follows. In Sec.~\ref{sec:related}, we review the related works. Sec.~\ref{sec:data} details the data collection and pre-processing steps. In Sec.~\ref{sec:global_perspective}, we examine the diffusion of news videos from a global perspective, followed by a more detailed analysis of US and German videos in Sec.~\ref{sec:fine_grained}.  Sec.~\ref{sec:results} discusses the problem formulation, modeling approaches, and evaluation of the models. Sec.~\ref{sec:conclusion} summarizes the findings.

\section{Related Works}\label{sec:related}

Given the breadth of literature on dissemination of different news modalities, here we confine the scope of reviewed literature to studies focusing on video news. For a more general review the reader is referred to~\cite{ORELLANARODRIGUEZ201874, E220362}.



\paragraph{\textbf{Diffusion of YouTube Videos:}}
A number of studies have investigated the diffusion patterns of YouTube videos, providing valuable insights into the factors influencing their popularity and engagement. Figueiredo et al. \cite{Figueiredo} emphasized the impact of popularity dynamics on social media, offering important considerations for content creators and advertisers. Trzcinski et al. \cite{trzcinski2017predicting} presented a model that utilizes Twitter trends to predict the popularity of YouTube videos, highlighting the interplay between different social media platforms. Morcillo et al. \cite{morcillo2016typologies} analyzed popular science web videos, uncovering the unique features that contribute to their success. In \cite{brodersen2012youtube}, the authors analyzed the consumption from the geospatial perspective, demonstrating that despite the global nature of the platform, the consumption appears constrained by the geographic locality of interest. \looseness-1

Basch et al. \cite{Basch2021COVIDSampling} assessed the effectiveness of informational videos about COVID-19 vaccination based on YouTube view counts and sorting algorithms, demonstrating how view metrics can serve as proxies for content popularity. On the other hand, Han et al. \cite{han2022assessment} conducted a cross-sectional study on videos from China related to lung nodules, offering insights into how specific health-related content is represented and received on YouTube. Along these lines, Bai et al. \cite{bai2022quality} studied the quality of internet videos related to pediatric urology in Mainland China, contributing to the understanding of information dissemination through platforms like TikTok. \looseness-2

Munaro et al.\cite{munaro2021engage} identified features that influence digital consumer engagement with YouTube videos, such as language elements, linguistic style, subjectivity, emotion valence, and video category, which significantly affect views, likes, and comments. 

In \cite{figueiredo2011tube}, Figueiredo et al. investigated popularity growth patterns using three datasets, highlighting how different types of referrers affect user attraction to videos. Lee et al. \cite{lee2022whose} studied the collective attention and diffusion of videos on controversial political topics, showing differences in the attention span of left-leaning versus right-leaning videos on YouTube and Twitter. Zeni et al. \cite{zeni2016understanding} investigated the reasons behind the virality of videos, examining how the popularity of tagged videos evolves over time and the existence of recurrent patterns in content popularity dynamics. The findings suggest that capturing these phenomenaargue requires a multi-scale, multi-level model. Abisheva et al. \cite{abisheva2014watches} combined Twitter and YouTube data to analyze the demographics, location, interests, and behaviors of users who share YouTube links, identifying trends based on the timing of sharing events. Anand et al. \cite{Anand2022} modeled the growth patterns of YouTube videos based on view count data and identified a multimodal growth pattern driven by initial subscribers and subsequent viral sharing.

\paragraph{\textbf{Popularity Prediction of \yt Videos:}}
A considerable body of literature has been published on the prediction of YouTube videos' popularity measures. For instance, Mekouar et al. \cite{mekouar2017popularity} examined machine learning-based approaches for predicting web content popularity, proposing a model that combines engineered features and word embeddings to predict video popularity. Similarly, Yu et al. \cite{yu2014twitter} proposed a method to predict YouTube video view-count increases driven by Twitter, achieving high precision by leveraging Twitter-derived features to predict early and sudden popularity jumps. Tafesse \cite{Tafesse2020} analyzed over 4000 YouTube trending videos and found that while lengthy and information-dense video titles negatively impact views, intense negative emotional sentiment in titles and information-dense descriptions positively influence views. An inverted U-shaped relationship was also observed between the number of video tags and views, with up to 17 tags increasing views, yet beyond that the opposite effect is observed. Yang et al. \cite{yang2010predicting} proposed a model to predict users' intentions for sharing YouTube videos, considering gender differences in behavior, where female users are influenced by perceived usefulness and social norms, while male users are more influenced by interpersonal norms. \looseness-2

Nisa et al. \cite{nisa2021optimizing} employed XGBoost to classify popular videos on YouTube and additionally evaluated the model performance on a dataset of unpopular videos. Ma et al. \cite{ma2017larm} studied long-term video popularity in complex YouTube networks, proposing LARM, a lifetime-aware regression model that outperformed several non-trivial baselines on two YouTube datasets with different observation intervals. Taking a different approach, Wongsuparatkul et al. \cite{wongsuparatkul2020view} proposed a multivariate linear model, which utilizes clustering and kNN to predict the short-term view count of YouTube videos. The model was compared against the best baseline model from the literature, demonstrating considerable performance gains on the 30th-day view count prediction. Meanwhile, Batta et al. \cite{batta2022predicting} defined YouTube video popularity as a combination of viewer engagement features such as view, like, and subscriber counts and achieved over 90\% accuracy in predicting the new popularity measure with a Random Forest Classifier. Further, Manikandan et al. \cite{manikandan2022prediction} analyzed trending videos on YouTube from five countries to identify predictive factors of view counts. The analysis revealed comment count, country, published date and video category being the most informative ones. Additionally, the authors compare the performance of various machine learning models in forecasting view counts based on video metadata and information regarding the trending date and country. In Aljamea et al. \cite{Aljamea}, the authors use a comprehensive dataset of YouTube videos for next-hour video popularity prediction, leveraging various features ranging from video and channel metadata to sentiment scores of the comments. The results are compared across different traditional and statistical machine learning models. Lastly, Pinto et al. \cite{pinto2013using} incorporated early view patterns to predict the popularity of YouTube videos using various models, validating the reduction in relative squared error on trending and random sample of videos. Put together, these studies illustrate the multifaceted nature of video popularity prediction and the wide array of correlates, ranging from textual content and emotional sentiment to social media activity and viewer engagement. \looseness-2

\section{Datasets}\label{sec:data}

To investigate early-stage diffusion patterns and dispersion rates of YouTube news videos, we compiled two distinct datasets, referred to as Dataset A and Dataset B, containing over 55,000 videos in total. Dataset A spans 75 countries and six continents, providing a global perspective, while Dataset B focuses specifically on the United States and Germany. These datasets were collected using different methods, each tailored to specific regional focuses. Table~\ref{tab:channel_stats} provides an overall comparison between them, showcasing key attributes like country coverage and channel statistics.




\begin{table*}[!h]
    \center
        \caption{Metadata details on the channels in the collected data (both Dataset A and B). The channel age is calculated based on the creation year and the year of data collection.}
    \resizebox{\textwidth}{!}{%
    \begin{tabular}{@{\extracolsep{5pt}}lccccc}
        \\[-1.8ex]\hline
        \hline \\[-1.8ex]
        &  \textit{Dataset A}  & \multicolumn{2}{c}{\textit{Dataset B}}\\
        \cr \cline{2-4}
         &   & US & Germany \\
         \cline{2-4} \\[-1.8ex]
        Number of videos retrieved & 48,406 &	5,568 & 5,023 \\
        Average no. of videos per channel retrieved & 77 & 	152 & 146\\
        Average no. of videos per channel  & 12,016 & 	30,202 & 12,490\\
        Average no. of subscribers & 1,220,372 & 3,829,200 & 1,603,406 \\
        Average channel age in years & 9.73	& 12.14 & 13.55\\
        Average channel views & 514,024,146 & 2,749,927,688 & 479,187,157  \\
        \hline \\[-1.8ex]
        Number of channels & 624 & 38 & 33  \\
        \hline \\[-1.8ex]
    \end{tabular}%
    }
    \label{tab:channel_stats}%
\end{table*}

\paragraph{\textbf{Dataset A}:} The first dataset aims at quantifying and analyzing half-life across different countries within the News and Politics category on \yt. First, a list of channels was obtained from HypeAuditor~\cite{hypeauditor}. Specifically, for each country available, up to 50 most subscribed and most commented channels were obtained within the category of interest. Then, those channels that were not monitored by ViewStats~\cite{viewstats} were filtered out, leaving 1,267 channels in the dataset. For the remaining channels, we retrieved a random sample of videos published within the previous 3 months, and scraped hourly views of the first 24 hours collected by ViewStats. YouTube Data API v3 was employed to extract video metadata. The resulting dataset features 81 countries and 51,073 videos. We further exclude any videos from the analysis if the hourly views are not available for at least 12 of the 24 hours, leaving 48,406 across 624 channels. \looseness-2

\paragraph{\textbf{Dataset B}:} The motivation for the second dataset is to get a more granular data enabling the analysis of diffusion dynamic patterns across different types of news channels, and assess the predictive capacity of a set of features that include video and channel metadata. To this end, we compiled a list of over 70 channels from two regions: Germany and the US. Furthermore, we made sure to include both mainstream and alternative news channels.  We collected a 24-hour time span of videos posted by the above channels from August 15 to September 30. The data collection comprises of two processes running simultaneously. The first process ensures capturing a newly posted video from the identified channels, and adds to the queue for the second process. The latter, extracts metadata about the video and monitors the views for the following 24 hours. The first data collection process ensures the prompt capture of newly posted videos from the identified channels, typically within an average time frame of 3.94 minutes after publication, facilitating real-time analysis of emerging content. The jobs for collecting the metadata information and temporal information for \yt channels and videos are developed with parallel task scheduler that uses the Python's built-in multiprocessing module and runs every 5 minutes. Once a 24-hour timeframe is created for a particular video, the video is moved to another queue which maintains the set of completed videos. The final dataset contains 10,591 videos from 71 channels.

~\\
Both datasets contain three groups of features related to channel metadata (e.g., number of subscribers, total views, and joined year), video metadata (e.g., title, length, publication date, and time), and temporal information regarding videos' likes and views. Due to page limitations, the detailed descriptions of features are deferred to Table~\ref{tab:all-features} in Sec.\ref{sec:featdetails}.

\subsection{Data Pre-processing}
Below, we describe the pre-processing and feature engineering steps that were applied on each dataset and highlight the nuances. 

\paragraph{\textbf{Pre-processing of Dataset A}:} To quantify the half--life, videos with more than one missing value during the first 12 hours were dropped. The median percentage change from half--life values to the actual view-count of the hour when the half--life is reached is 9.56, which further verifies that half-life for these videos is reached within the above-mentioned time span. As a next step, GPT--4o model was utilized to label the titles into the following categories: politics,  economy/finance, entertainment, war/crime/conflict and other. \looseness-2

\paragraph{\textbf{Pre-processing of Dataset B}:} In the initial stage of data processing, videos lacking complete temporal data over the 24-hour period were excluded from the dataset. Additionally, videos with more than two consecutively missing data points in the time-series were also excluded, enhancing the reliability of our analysis. The remaining missing values were replaced by the average of the adjacent non-missing values. As with \textit{Dataset A}, we employ GPT--4o to annotate the titles, however we use more granular categories. \looseness-2

\paragraph{\textbf{Feature Engineering}:} 

To enhance our analysis, we employed GPT-4o for automated feature extraction from video titles (please see Table~\ref{lst:prompt} for the prompt). Specifically, the titles were annotated with a range of linguistic and semantic features. For each title the following features were extracted:

\begin{itemize}
    \item Sentiment: Classified as either positive, neutral, or negative, providing insight into the emotional tone.
    \item Subjectivity: A binary indicator of whether the title is objective or subjective, reflecting the degree of factual versus opinion-based content.
    \item  Emotion: Categorized emotions (e.g., neutral, happy, sad) to capture the underlying emotional state.
    \item Number of Named Entities: Number of detected entities such as persons, locations, or organizations. 
    \item Verb Tense: Identified the dominant tense (e.g., present, past) used in the title to infer temporal relevance.
    \item Urgency: A qualitative assessment of urgency, rated as low, medium, or high.
    \item Other Linguistic Properties: Included binary indicators for the presence of emojis, whether the title is a question.
\end{itemize}

\section{Half-life of YouTube Videos: A Global Perspective}\label{sec:global_perspective}

We assess the rate of video views diffusion by calculating the time required for videos to reach their 24-hour half-life, representing the point at which half of the total anticipated views within a 24-hour timeframe have been achieved. In Table~\ref{tab:half_life_quantiles_world}, we show the distribution of half-lives in hours in our sample within Dataset A. \looseness-2

\begin{table}[!b]
    \centering
    \begin{minipage}{0.45\textwidth}
        \centering
        \caption{Quantiles of half-lives in hours (Dataset A).}
        \label{tab:half_life_quantiles_world}
        \begin{tabular}{lccccc}
            \toprule
            \textbf{Quantiles} & 10\% & 25\% & 50\% & 75\% & 90\% \\
            \midrule
            Half-Life & 3 & 4 & 6 & 7 & 10 \\
            \bottomrule
        \end{tabular}
    \end{minipage}%
    \hfill
    \begin{minipage}{0.47\textwidth}
        \centering
        \caption{Quantiles of half-lives converted to hours (Dataset B).}
        \label{tab:half_life_quantiles}
        \begin{tabular}{lccccc}
            \toprule
            \textbf{Quantiles} & 10\% & 25\% & 50\% & 75\% & 90\% \\
            \midrule
            US & 3.17 & 4.17 & 5.58 & 7.5 & 9.67 \\
            Germany & 3.75 & 4.67 & 5.92 & 7.92 & 10.58 \\
            \bottomrule
        \end{tabular}
    \end{minipage}
\end{table}

\begin{figure}[!htbp]
    \centering
    \includegraphics[scale=0.42]{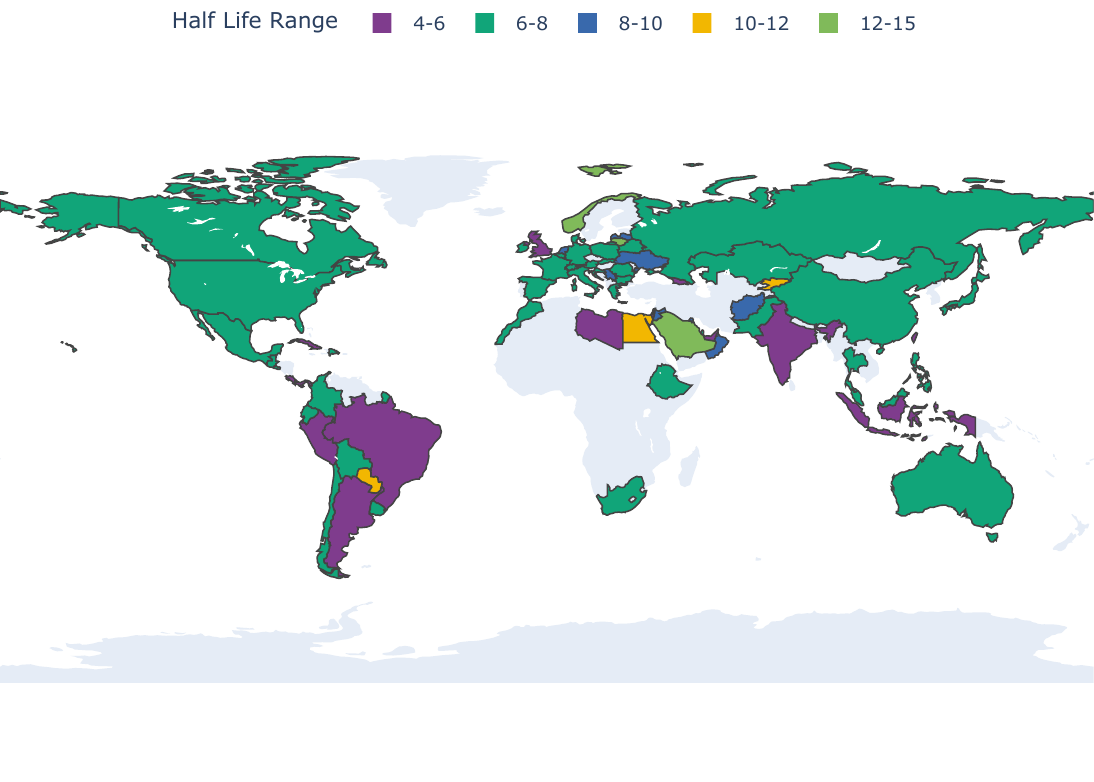}
    \caption{Average half-life in hours per country binned into 5 categories in Dataset A. Countries with no data are colored gray.}
    \label{fig:half_life_map}
\end{figure}

It can be observed that while a considerable portion (25\%) of videos achieved their half-lives relatively quickly (within 4 hrs.), there is a large variation. We further look into the country-wise average half-lives, which is illustrated in Fig.~\ref*{fig:half_life_map}. Differences in half-life averages across countries can be attributed to various cultural, societal, and economic factors. For example, countries with more stable political environments might see longer engagement with news videos as the content remains relevant. On the other hand, countries experiencing rapid changes or high news turnover might have shorter half-life values as new events quickly overshadow older news. As seen from Fig.~\ref*{fig:half_life_map}, country-wise average half-lives vary significantly, from 4 to 15 hrs. Interestingly, the longest country-wise averages (ranging from 10 to 15 hours) were observed in the Middle East and Egypt. \looseness-2

\section{Fine-grained Analysis of News Videos in Germany and US}\label{sec:fine_grained}

In Table \ref{tab:half_life_quantiles}, we present the quantiles for half-life durations in minutes across our \textit{Dataset B}, which reveals differences in half-lives across the two countries.  Specifically, videos from the US news channels reach their half-lives earlier compared to German videos (one-sided t-test, $p<0.001$), suggesting that videos from the US channels gain traction more quickly and may posses a higher level of audience engagement or a faster spread of content. \looseness-2

\begin{figure*}[!h]
    \centering
    \includegraphics[width=1.0\textwidth]{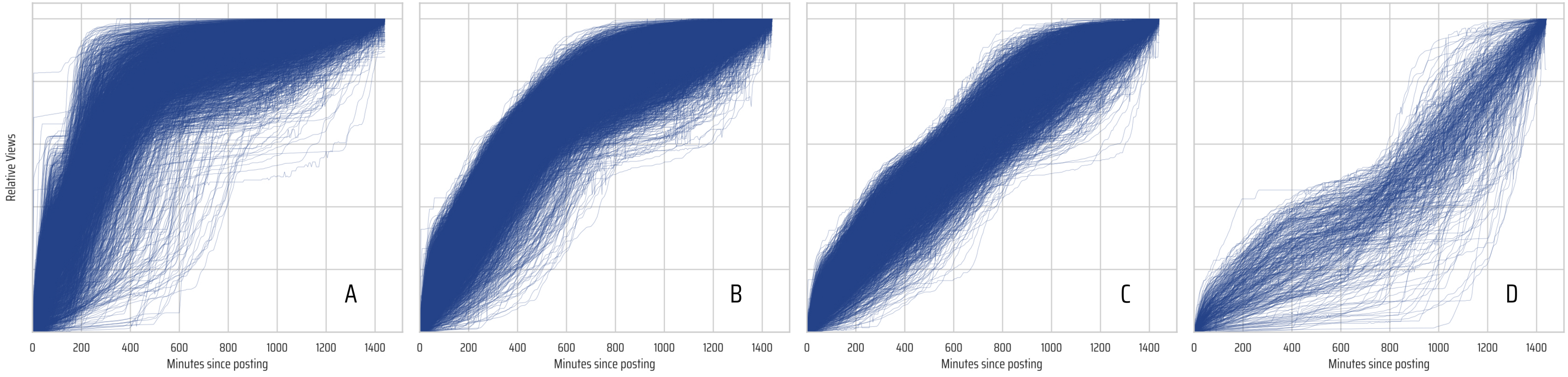}
    \caption{k-Shape clustering results with the best $k$ depicting four different patterns of diffusion.}
    \label{fig:clusters}
\end{figure*}

\paragraph{\textbf{Diffusion Patterns:}}
In Fig.~\ref{fig:clusters}, we present the views over time (in minutes) for all the videos in Dataset B, identifying various diffusion patterns of YouTube news videos. Four different diffusion patterns were observed:
\begin{itemize}
    \item \textit{Sigmoid growth (Cluster A):} This type of growth occurs in videos that rapidly achieve maximum peak visibility and then taper off in the end. The videos resemble viral news content that gets recommended heavily at first. An example could be a news report on an emerging health concern where there is a rapid increase as the story gains traction and growth starts to slowly approach saturation. 
    \item \textit{Logarithmic growth (Cluster B):} This pattern usually occurs in videos where views continue to increase but at a decreasing rate over time. An example could be breaking news about a major global event where the video gets a massive initial surge as users seek the initial information. However, as the event becomes less current, it continues to gain views at a decreasing rate.
    \item \textit{Linear growth (Cluster C):} This diffusion pattern usually occurs in videos with steady growth over time, which is more visible in consistently popular news channels with a stable subscriber base. An example could be a YouTube news channel posting weekly news recaps that gain views in a linear pattern as regular viewers of the channel tune in for their weekly updates. 
    \item \textit{Stepped growth (Cluster D):} This cluster could be best characterized by news videos with periods of sudden increases followed by plateaus. An example could be an investigative report on a developing story. This stepped growth is also an indication of possible artificial inflation of views (see e.g., \cite{10.1145/3477300}) or bot-related influence. 
 
    
\end{itemize}

\vspace*{-20pt}
\section{The Latency of Half-life: Predictive Models and Factors}
\label{sec:results}

To complement the descriptive analysis presented above, in this section, we examine the latency of \yt news videos through predictive analysis, evaluate and compare the performance of various Machine Learning and Deep Learning models, and investigate the factors that influence the predictions through Shapley Additive Explanations (SHAP)~\cite{lundberg2017unified} method.\looseness-2

\subsection{Prediction Problem and Models}\label{sec:modelling}

\paragraph{\textbf{Problem Statement}:} We study the problem of predicting the latency of \yt news videos' half-lives with the goal of identifying the most influential features that drive the distinction between ``early'' and ``late'' half-lives. To this end, we cast the problem as a classification task, with the target variable derived from percentile binning based on half-life values. Recall that the two datasets introduced in this work were constructed primarily for the purposes of a descriptive analysis. In order to cover a representative random sample of videos, no manual control over video selection was applied. Dataset A was more suitable for a classification task due to its global scope and balance in terms of country representation. Hence, it was selected to train the predictive models. Following similar analysis approach as in\cite{10.1007/978-3-030-64583-0_24}, we further divide the dataset into three quantiles: the bottom 30\% representing the earliest half-life, the middle 40\%, and the top 30\% representing the latest half-life. The factors responsible for the “middle” performance are likely the ones with negligible impact, hence inferring them might obscure the results. On the other hand, omitting a large group of subjects could lead to the loss of pertinent data. Thus, to steer the classification towards distinguishing early vs late videos (i.e., the interesting cases), we retained only the top and bottom 30\% (resulting in a binary classification task of ''early vs late''), ensuring a solid number of videos is included while retaining a sizable gap between the two classes. To formalize, let $X \in \mathbb{R}^{m}$ denote the vector of input features, where $m$ stands for the number of considered predictors. Here, $m=25$ and includes both numerical and categorical predictors extracted from Dataset A (please see Table \ref{tab:all-features} in Sec.~\ref{sec:featdetails}). Then, the problem can be formulated as learning a mapping of the form $X \rightarrow Y \in \{0,1\}$, where the prediction target $Y$ denotes the latency of half-lives values with $0$ corresponding to early reaching videos (i.e., the bottom $30\%$ percentile). \looseness-2

\paragraph{\textbf{Predictive models}:}

The predictive capacity of the considered 25 features (please see Table~\ref{tab:all-features} in Sec.~\ref{sec:featdetails}) was evaluated using two categories of models: statistical Machine Learning models and Deep Learning models. From the former category, we employed Logistic Regression, Support Vector Classifier (SVC), and XGBoost. From the latter, we considered a standard Deep Neural Network (DNN), the FT-Transformer~\cite{gorishniy2021revisitingFT}, and the Kolmogorov-Arnold Neural Network (KAN)~\cite{liu2024kan}. Additionally, we employ a naive classifier, referred to as Baseline, where the predicted category is derived from the channel's average half-life. Specifically, if a channel's average half-life is categorized as early or late, all videos from that channel are assigned to the corresponding category.\looseness-2

The dataset was split into 80\% for training and 20\% for testing, and the reported results are based on the testing set. The hyper-parameters of all the models were optimized through Grid Search.\looseness-2

\begin{table}[h]
\centering
\caption{Performance of candidate models on the Test dataset. Best scores are highlighted in bold.}
\small 
\setlength{\tabcolsep}{3pt} 
\renewcommand{\arraystretch}{0.9} 
\resizebox{0.85\columnwidth}{!}{%
\begin{tabular}{lccccc} 
\toprule \textbf{Model} & \textbf{Accuracy} & \textbf{Precision} & \textbf{Recall} & \textbf{F1-Score} & \textbf{ROC AUC}\\ 
\midrule 
\textit{Baseline} & 71.35 & 75.23 & 71.35 & 71.06 & 72.96\\ 
\textit{Logistic Regression} & 74.59 & 74.60 & 74.59 & 74.37 & 73.78\\ 
\textit{SVC} & 76.33 & 76.35 & 76.33 & 76.16 & 75.60 \\ 
\textit{XGBoost} & \textbf{82.51} & \textbf{82.55} & \textbf{82.51} & \textbf{82.42} & \textbf{81.97}\\ 
\textit{DNN} & 76.69 & 76.64 &76.69 & 76.61 & 76.17\\ 
\textit{FTTransformer} & 78.20 & 78.24 & 78.20 & 78.22 & 78.05 \\ 
\textit{KAN} & 77.36 & 77.33 & 77.36 & 77.26 & 76.79\\ 
\bottomrule 
\end{tabular}%
}
\label{tab:results}
\end{table}

The performance comparison across models, reported in Table \ref{tab:results}, highlights XGBoost as the top performer, significantly outperforming other models with an accuracy of 82.51\% and the highest scores across all the considered metrics. This indicates its superior ability to correctly classify videos into early and late half-life categories. In contrast, traditional models like Logistic Regression and SVC demonstrated moderate improvements over the baseline, with SVC achieving slightly better accuracy (76.33\%) and ROC AUC (75.60\%) than Logistic Regression (74.59\% accuracy). Deep learning models such as the DNN and FTTransformer offered a slight boost over SVC, but their performance still lagged behind XGBoost. Notably, the FTTransformer shows stronger ROC AUC (78.05\%), suggesting good discriminatory power, while KAN performs similarly to the other models yet fails to match the robustness of XGBoost. Overall, XGBoost’s dominance indicates its suitability for this task, balancing both predictive accuracy and class separability. \looseness-2

\subsection{Predictive Factors} 

\begin{figure}[!ht]
    \centering
    \includegraphics[width=0.6\textwidth]{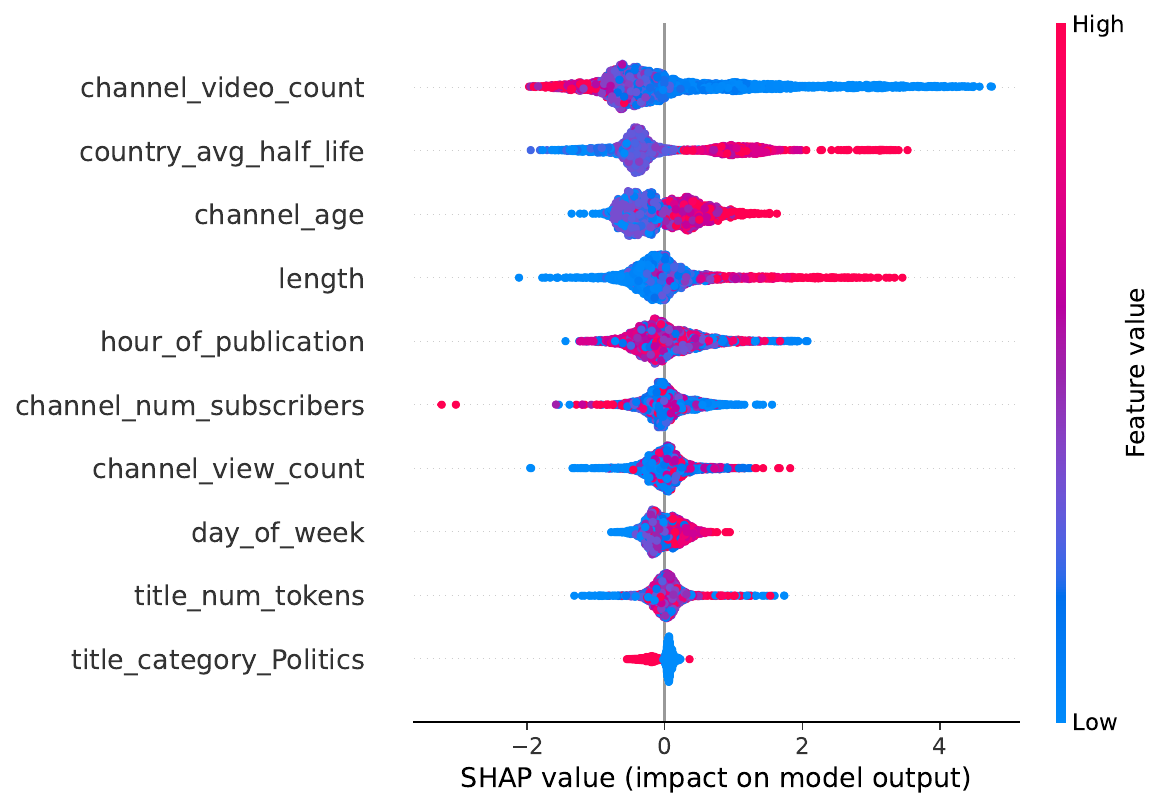}
    \caption{Impact of Top 10 features on XGBoost model. }
    \label{fig:shap}
\end{figure}

As the preceding section revealed, the highest accuracy of predictions was attained with XGBoost. However, XGBoost is a black-box ensemble-tree model, restricting the capability to directly quantify the impact of individual features on the latency of \yt videos' half-life. To address this limitation, we employ SHAP. Fig.~\ref{fig:shap} depicts the impact of top 10 features on the model's predictions. The most influential feature is \textit{title\_category\_Politics}, suggesting that videos categorized as political exhibit distinct half-life patterns. Temporal aspects, such as \textit{day\_of\_week and hour\_of\_publication}, underscore the importance of posting time in shaping a video's lifespan. The length of the video has a further impact, suggesting that shorter videos tend to attract the viewers more quickly. The more videos the channel uploads, the greater its potential impact on the model’s predictions, as suggested by the \textit{channel\_video\_count} feature. A higher number of uploads may indicate an active channel with consistent engagement, which can influence the half-life of its content. However, channel\_age, does not necessarily translate to earlier half-lives. This suggests that mere longevity on YouTube does not inherently lead to faster content decay; rather, engagement-driven factors such as frequent uploads and audience interaction play a more critical role in determining a video's lifespan.  \looseness-2

\section{Conclusion}\label{sec:conclusion}

This study presented an in-depth analysis of the early-stage diffusion dynamics of YouTube news videos, focusing on their 24-hour half-life. Two extensive datasets were compiled and analyzed. We identified four distinct diffusion patterns: sigmoid, logarithmic, linear, and stepped growth. These patterns reflect variations in how users engage with news videos across different contexts. Additionally, we analyzed the latency of half-lives using predictive modeling, where an ensemble XGBoost classifier emerged as the most effective model, achieving an average F1-score of 82\%. The findings from our Explainable AI (XAI) analysis revealed key predictive factors, such as the channel’s video posting frequency, country-level engagement, and the video's duration. Beyond academic interest, these insights hold practical implications for content strategists, platform moderators, and algorithm designers. For instance, early identification of fast-diffusing videos can aid in timely promotion or moderation, while media organizations can adjust upload timing and content framing to better align with expected diffusion patterns. \looseness-2


\bibliographystyle{splncs04}
\bibliography{references}

\newpage
\appendix

\pagenumbering{gobble}
\raggedbottom
\section*{Supplementary Materials}
\section{Feature Details}\label{sec:featdetails}

\vspace{-0.5cm}

\begin{table}
\centering
\caption{Description of features in the datasets. The features marked with asterisk comprise the set of 25 predictors used for the classification of half-life latency.}
\resizebox{\textwidth}{!}{%
\begin{tabular}{ll}

\toprule
\toprule
\textbf{Feature Name} & \textbf{Description} \\
\midrule
\multicolumn{2}{c}{\textit{Channel Metadata (scraped data)}} \\
\cmidrule(l{14em}r{14em}){1-2}
channel\_id & Unique identifier assigned by YouTube. \\
channel\_view\_count\textsuperscript{*} & Total views accumulated by the channel. \\
channel\_video\_count\textsuperscript{*} & Number of videos uploaded by the channel. \\
channel\_num\_subscribers\textsuperscript{*} & Number of subscribers on the channel. \\
joined\_year & Year the channel was created. \\
\cmidrule(l{14em}r{14em}){1-2}

\multicolumn{2}{c}{\textit{Video Metadata (scraped data)}} \\
\cmidrule(l{14em}r{14em}){1-2}
video\_id & Unique identifier of the video. \\
title & Title of the video. \\
length\textsuperscript{*} & Duration of the video in seconds. \\
is\_age\_restricted & Indicates if the video is age-restricted. \\
thumbnail\_url & URL of the video's thumbnail. \\
channel\_id & Identifier of the channel that uploaded the video. \\
published\_datetime & Date and time when the video was published. \\
num\_comments & Number of comments on the video. \\
\cmidrule(l{14em}r{14em}){1-2}
\multicolumn{2}{c}{\textit{Engineered Features}} \\
\cmidrule(l{14em}r{14em}){1-2}
sentiment\textsuperscript{*} & Sentiment of the video title (-1: Negative, 0: Neutral, 1: Positive). \\
subjectivity\textsuperscript{*} & Indicates whether the title is subjective or objective. \\
has\_named\_entities\textsuperscript{*} & Whether the title contains named entities (1: Yes, 0: No). \\
urgency\textsuperscript{*} & Perceived urgency of the title (1: Low, 2: Medium, 3: High). \\
is\_emotional\textsuperscript{*} & Whether the title contains emotional content (1: Yes, 0: No). \\
has\_emojis\textsuperscript{*} & Indicates if emojis are present in the title. \\
is\_question\textsuperscript{*} & Whether the title is phrased as a question. \\
verb\_tense\textsuperscript{*} & Tense of the verb in the title (1: Past, 2: Present, 3: Future). \\
day\_of\_week\textsuperscript{*} & Day of the week the video was published. \\
hour\_of\_publication\textsuperscript{*} & Hour of the day the video was published. \\
title\_num\_tokens\textsuperscript{*} & Number of tokens in the title. \\
channel\_age\textsuperscript{*} & Age of the channel in years. \\
title\_category\_Conflict\textsuperscript{*} & Binary feature indicating if the title is related to conflict. \\
title\_category\_Economy\textsuperscript{*} & Binary feature indicating if the title is related to economy. \\
title\_category\_Health\_and\_Safety\textsuperscript{*} & Binary feature indicating if the title is related to health and safety. \\
title\_category\_Other\textsuperscript{*} & Binary feature indicating if the title falls under other categories. \\
title\_category\_Politics\textsuperscript{*} & Binary feature indicating if the title is related to politics. \\
title\_category\_Science/Tech\textsuperscript{*} & Binary feature indicating if the title is related to science and technology. \\
title\_category\_Society\textsuperscript{*} & Binary feature indicating if the title is related to society. \\
title\_category\_Sports\textsuperscript{*} & Binary feature indicating if the title is related to sports. \\
country\_avg\_half\_life\textsuperscript{*} & Average half-life of videos in that country. \\
\bottomrule
\bottomrule
\end{tabular}
}
\label{tab:all-features}
\end{table}

\section{Details of Prompt Engineering}
\label{sec:promptdetails}

\begin{minipage}{0.95\textwidth}
\captionof{table}{Prompt Engineering Instructions}  
\label{lst:prompt}  

\begin{minipage}{\textwidth} 
\small
\begin{lstlisting}

You are a research assistant specializing in analyzing video titles across different languages. 
Your task is to extract detailed information from the given video title while ensuring the following:

1. Title Cleaning: Remove channel names, dates, and extraneous words that do not contribute to the main meaning.

2. Sentiment Analysis: Determine if the title expresses a positive, negative, or neutral sentiment.

3. Subjectivity: Identify whether the title is subjective, objective, or neutral in tone.

4. Emotion Detection: Categorize the dominant emotion as happy, sad, angry, fear, surprise, or neutral.

5. Named Entities: Identify and extract entities related to persons, organizations, and countries if they are present.

6. Urgency: Classify the urgency level as high, medium, or low based on wording and context.

7. Verb Tense: Determine whether the title refers to an event in the past, present, or future.

8. Other Features:
   - Check if the title contains emojis.
   - Detect if the title is phrased as a question.
   - Identify if a date is explicitly mentioned.
   - Detect the language of the title.
   - Provide an English translation if the title is not in English.
   - Classify the topic into one of the following: politics, economy, entertainment, crime and investigation, war and conflict, 
     religion, environment, sports, social, bulletin news, or other.

Input:
""

Output Format:
Provide the extracted information in a structured JSON format adhering to the schema of TitleInfo.

Ensure that the results are accurate, structured, and concise.

\end{lstlisting}
\end{minipage}
\end{minipage}

\end{document}